\documentclass[pdflatex,sn-mathphys-num]{sn-jnl}

\usepackage{graphicx}%
\usepackage{multirow}%
\usepackage{amsmath,amssymb,amsfonts}%
\usepackage{amsthm}%
\usepackage{mathrsfs}%
\usepackage[title]{appendix}%
\usepackage{xcolor}%
\usepackage{textcomp}%
\usepackage{manyfoot}%
\usepackage{booktabs}%
\usepackage{algorithm}%
\usepackage{algorithmicx}%
\usepackage{algpseudocode}%
\usepackage{listings}%
\usepackage{algorithm}%
\usepackage{tikz}
\usetikzlibrary{positioning}

\theoremstyle{thmstyleone}%
\theoremstyle{thmstyletwo}%

\theoremstyle{thmstylethree}%

\begin{document}

\title[Article Title]{POST: Photonic Swin Transformer for Automated and Efficient Prediction of PCSEL}


\author[1,2]{\fnm{Qi} \sur{Xin}}\email{qixin@link.cuhk.edu.cn}
\equalcont{These authors contributed equally to this work.}

\author[1,2]{\fnm{Hai} \sur{Huang}}\email{haihuang@link.cuhk.edu.cn}
\equalcont{These authors contributed equally to this work.}

\author[1]{\fnm{Chenyu} \sur{Li}}\email{chenyuli@link.cuhk.edu.cn}

\author[3]{\fnm{Kewei} \sur{Shi}}\email{kewei.shi@connect.hku.hk}


\author*[1,2]{\fnm{Zhaoyu} \sur{Zhang}}\email{zhangzy@cuhk.edu.cn}

\affil[1]{School of Science and Engineering, The Chinese University of Hong Kong, Shenzhen, Guangdong 518172, China}
\affil[2]{Guangdong Key Laboratory of Optoelectronic Materials and Chips and Shenzhen Key Lab of Semiconductor Lasers, School of Science and Engineering, The Chinese University of Hong Kong, Shenzhen, Guangdong 518172, China}
\affil[3]{\orgname{The University of Hong Kong}, \orgaddress{\street{Pok Fu Lam Road}, \city{Hong Kong}, \country{China}}}


\abstract{This work designs a model named POST based on the Vision Transformer (ViT) approach. Across single, double, and even triple lattices, as well as various non-circular complex hole structures, POST enables prediction of multiple optical properties of photonic crystal layers in Photonic Crystal Surface Emitting Lasers (PCSELs) with high speed and accuracy, without requiring manual intervention, which serves as a comprehensive surrogate for the optical field simulation. In the predictions of Quality Factor (Q) and Surface-emitting Efficiency (SE) for PCSEL, the R-squared values reach 0.909 and 0.779, respectively. Additionally, it achieves nearly 5,000 predictions per second, significantly lowering simulation costs. The precision and speed of POST predictions lay a solid foundation for future ultra-complex model parameter tuning involving dozens of parameters. It can also swiftly meets designers' ad-hoc requirements for evaluating photonic crystal properties. The database used for training the POST model is derived from predictions of different photonic crystal structures using the Coupled-Wave Theory (CWT) model. This dataset will be made publicly available to foster interdisciplinary research advancements in materials science and computer science.}




\maketitle

\section{Introduction}\label{sec1}

\begin{figure}[htbp]
    \centering
    \includegraphics[width=0.9\linewidth, trim=6 4 6 6, clip]{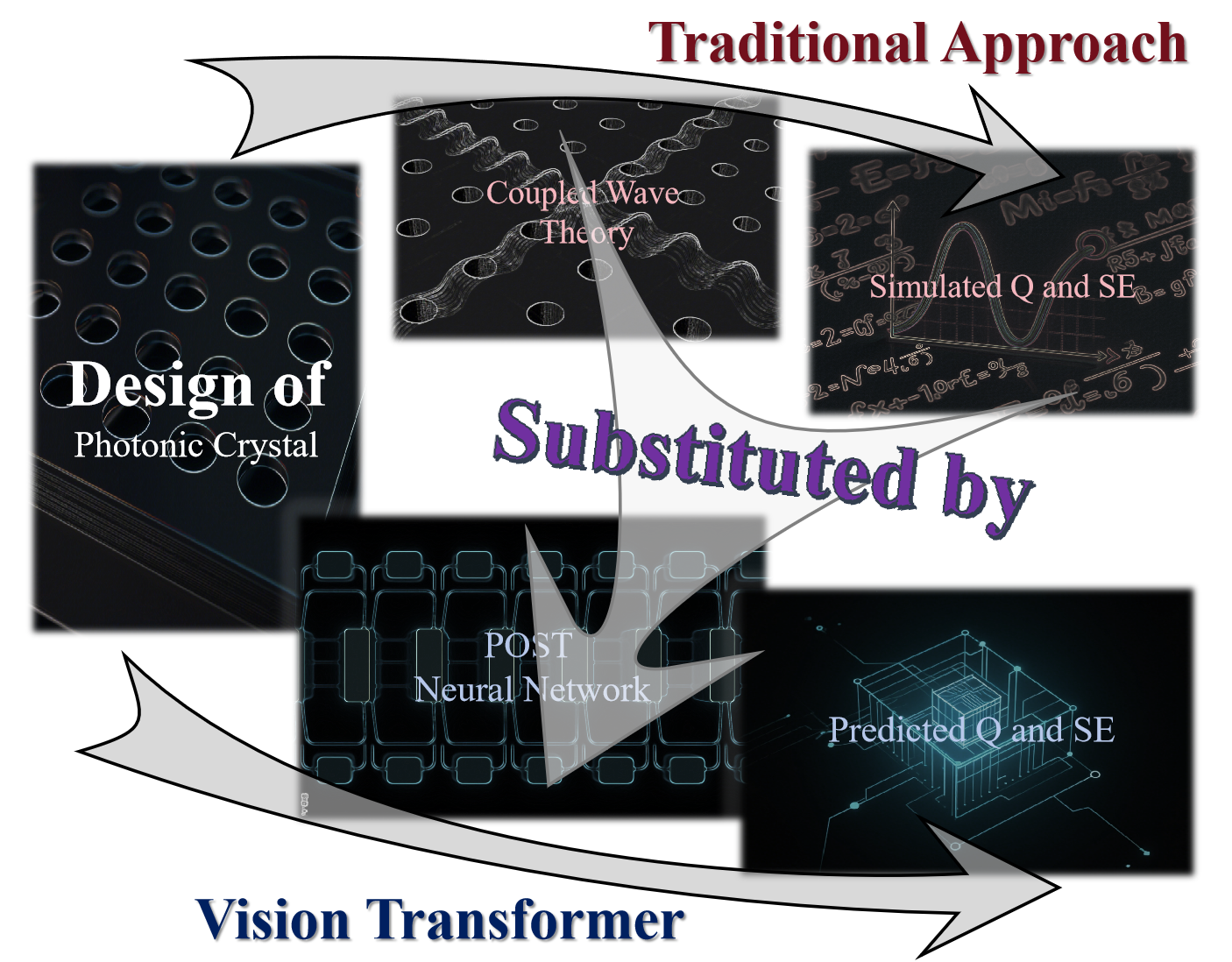}
    \caption{\textbf{Schematic overview.} This work replaces PCSEL's conventional simulation model with POST neural network prediction model, achieving a qualitative leap in the speed of design characterization.}
    \label{fig:abstract fig}
\end{figure}

Photonic Crystal Surface-Emitting Lasers (PCSELs) are a novel type of semiconductor lasers that achieve high power, high beam quality, and low divergence by applying a two-dimensional photonic crystal layer as optical resonant cavity\cite{yoshida_Highbrightness_2023, huang_Unveiling_2025}. The periodic modulation of the refractive index enables in-plane distributed feedback and vertical radiation, making PCSELs advantageous over traditional VCSELs in terms of scalability and output coherence\cite{ishimura_Proposal_2023, lang_Modelling_2023, noda_Highpower_2023}.
Conventional modeling techniques for EELs and VCSELs—such as Fabry–Pérot cavity analysis or 1D transfer matrix methods—are efficient for vertically layered structures but fail to capture the lateral periodicity and complex mode coupling in PCSELs. Numerical solvers like FDTD can handle these effects, but their high computational cost (hours per simulation) limits scalability\cite{lang_Modelling_2023, xu_Modeling_2024, sakai_Coupledwave_2010}.
CWT offers a more efficient alternative. By expanding electromagnetic fields into spatial harmonics, CWT accurately models in-plane diffraction and vertical radiation in photonic crystals\cite{liang_Threedimensional_2011, liang_Threedimensional_2012, liang_Threedimensional_2014}. It provides a good balance between physical accuracy and computational speed, making it especially suitable for large-scale PCSEL design\cite{yoshida_Highbrightness_2023}.

While CWT is significantly faster than full-wave solvers, it still requires several minutes per simulation for complex PCSEL unit cells— especially those with multi-lattice or irregular hole geometries:  Conducting 100,000 simulations will take nearly a year. 
Moreover, irregular geometries lead to the explosion of the number of design variables—potentially dozens or even over one hundred, which increases the volume of the simulation space exponentially. Optimization process on such space always requires millions or even billions of simulations. 

To overcome these challenges, an idea is to establish a database where a machine learning architecture can learn the physical principles of PCSEL's photonic crystal layers and makes efficient predictions, for example, within 0.001 seconds per sample. Then over 80 million predictions could be performed daily to satisfy the requirements of high-dimensional optimization. 

Vision Transformer (ViT) architecture is the go-to solution for this machine learning problem.
Inspired by the success of Transformers in natural language processing (NLP)~\cite{guo2024deepseekcoderlargelanguagemodel, glm2024chatglmfamilylargelanguage, devlin2018bert, sanh2020distilbertdistilledversionbert, brown2020languagemodelsfewshotlearners}, researchers began exploring their application to computer vision. ViT~\cite{dosovitskiy2021imageworth16x16words} proposed a purely Transformer-based architecture for image classification, challenging the dominance of convolutional neural networks (CNNs). ViT divides an image into fixed-size patches (e.g., $16\times16$), treats each patch as a token, and processes the resulting sequence using standard Transformer mechanisms, mirroring how sequences of words are handled in NLP.
ViT has demonstrated strong performance on image classification~\cite{dosovitskiy2021imageworth16x16words, chen2021crossvit}, object detection~\cite{carion2020endtoendobjectdetectiontransformers, li2022exploringplainvisiontransformer, fang2022unleashingvanillavisiontransformer, khoramdel2024yoloformeryoloshakeshand}, and semantic segmentation~\cite{zheng2021rethinkingsemanticsegmentationsequencetosequence, strudel2021, zhang2022segvit, zhang2023segvitv2}, and has been extensively improved through works such as DeiT~\cite{touvron2021training} and ConViT \cite{d2021convit} to enhance its training efficiency and accuracy. ViT offers advantages like lower computational cost and compatibility with Transformer-based optimization frameworks.

The Swin Transformer (SwinT)~\cite{liu2021swintransformerhierarchicalvision} is one of the ViT models. It computes self-attention within non-overlapping local windows, reducing complexity from quadratic to linear. The introduction of shifted windowing facilitates cross-window information flow and enhances local context modeling. These architectural innovations make SwinT a versatile backbone for a wide range of vision tasks and are key to its success in our proposed POST model.

Despite ViT's advantages, there has been limited application of ViT architectures to physical modeling tasks, particularly in photonics. Accurate modeling of the optical properties of complex PCSEL photonic crystal layers remains largely unexplored. This presents a significant research gap and a promising direction for applying ViT-based methodologies to advanced physical modeling and predictive tasks in photonics~\cite{chen2022povitvisiontransformermultiobjective, yan2025metasurfacevitgenericaimodel, sebastian2025prionvitprionsinspiredvisiontransformers}. 
The application of ViT in photonic crystal design faces challenges: ViT model has high data requirements, difficulties in visual reasoning and training stability issues, while design of photonic crystal requires simultaneously analyzing both the global and local physical properties effectively.

This work addresses these challenges by employing the Coupled-Wave Theory (CWT) model and the Photonic Swin Transformer (POST) model (The overview of this work is shown in Figure \ref{fig:abstract fig}). The POST model handles single-, double-, and triple-lattice configurations as well as arbitrarily shaped holes with smooth curved contours, supporting irregular geometries beyond circles or triangles. POST predicts multiple optical properties efficiently with high accuracy, achieving a simulation speed of about 5,000 samples per second and a training time under 1,000 seconds per epoch. POST is based on SwinT~\cite{liu2021swintransformerhierarchicalvision}, a state-of-the-art ViT architecture. The mean squared error of predictions is reduced by over 50\% compared to previous works, and it surpasses the prediction accuracy of existing methods with less than 20\% of the original dataset \cite{huang2025towards}.


\section{Results}

\subsection{Dataset Generation}

\subsubsection{Raw Data Acquisition via CWT}

\begin{table}[htbp]
        \caption{\textbf{Epitaxial structure of the PCSEL in this work}}
        \label{tab:epitaxy-structure}
        \centering
        \begin{tabular}{cccc}
            \hline
            Layer                    & Material      & Thickness ($\mu m$) & Refractive Index \\
            \hline
            Photonic Crystal         & p-GaAs/Air    & 0.35                & 3.4826/1         \\
            Waveguide                & p-GaAs        & 0.08                & 3.4826           \\
            Electron Blocking Layers & p-AlGaAs      & 0.025               & 3.2806           \\
            Active Region            & InGaAs/AlGaAs & 0.116               & 3.3944           \\
            n-cladding               & n-AlGaAs      & 2.11                & 3.2441           \\
            n-substrate              & n-GaAs        & -                   & 3.4826           \\
            \hline
        \end{tabular}
    \end{table}
The epitaxial structure listed in Table~\ref{tab:epitaxy-structure} serves as a representative baseline for PCSEL design. While the epitaxial configuration influences parameters such as the optical Green's function and the confinement factor of the photonic crystal ($\Gamma_{\mathrm{PhC}}$) within the CWT framework, its impact on the overall device behavior is secondary to the photonic crystal design. Therefore, the proposed methodology retains broad applicability and can be readily extended to alternative epitaxial stacks without significant modification.

To evaluate the optical performance of photonic crystal surface-emitting lasers (PCSELs), we adopt the CWT to model the interaction of fundamental waves within the photonic crystal (PhC) lattice. By considering four primary wave components propagating along orthogonal directions, a set of coupled partial differential equations is established to capture both diffraction feedback and radiation loss mechanisms\cite{liang_Threedimensional_2012a}.

Solving this model under appropriate boundary conditions for a finite-size square-lattice PhC allows us to compute the spatial distribution of optical fields, from which key performance metrics can be derived. Among these, two indicators are particularly crucial.
Surface-emitting Efficiency (SE) defined as the ratio between the surface-emitting optical power and the total stimulated emission power:
    \begin{equation}
        \mathrm{SE} = \frac{P_{\mathrm{surface}}}{P_{\mathrm{stim}}}
    \end{equation}
This ratio reflects how effectively the laser extracts optical energy through vertical radiation and serves as a direct metric for surface output optimization.
    
Quality Factor (Q) quantifies the ratio of stored optical energy to energy lost per oscillation cycle, expressed as:
    \begin{equation}
        Q = \frac{2\pi / a}{\alpha_{\mathrm{r}}}
    \end{equation}
where $a$ is the lattice constant and $\alpha_{\mathrm{r}}$ is the total radiation loss of the mode. A higher $Q$ indicates better optical confinement and lower lasing threshold.

\subsubsection{Data Preprocessing}

\begin{figure}[htbp]
    \centering
    \includegraphics[width=0.9\linewidth, trim=2 2 2 2, clip]{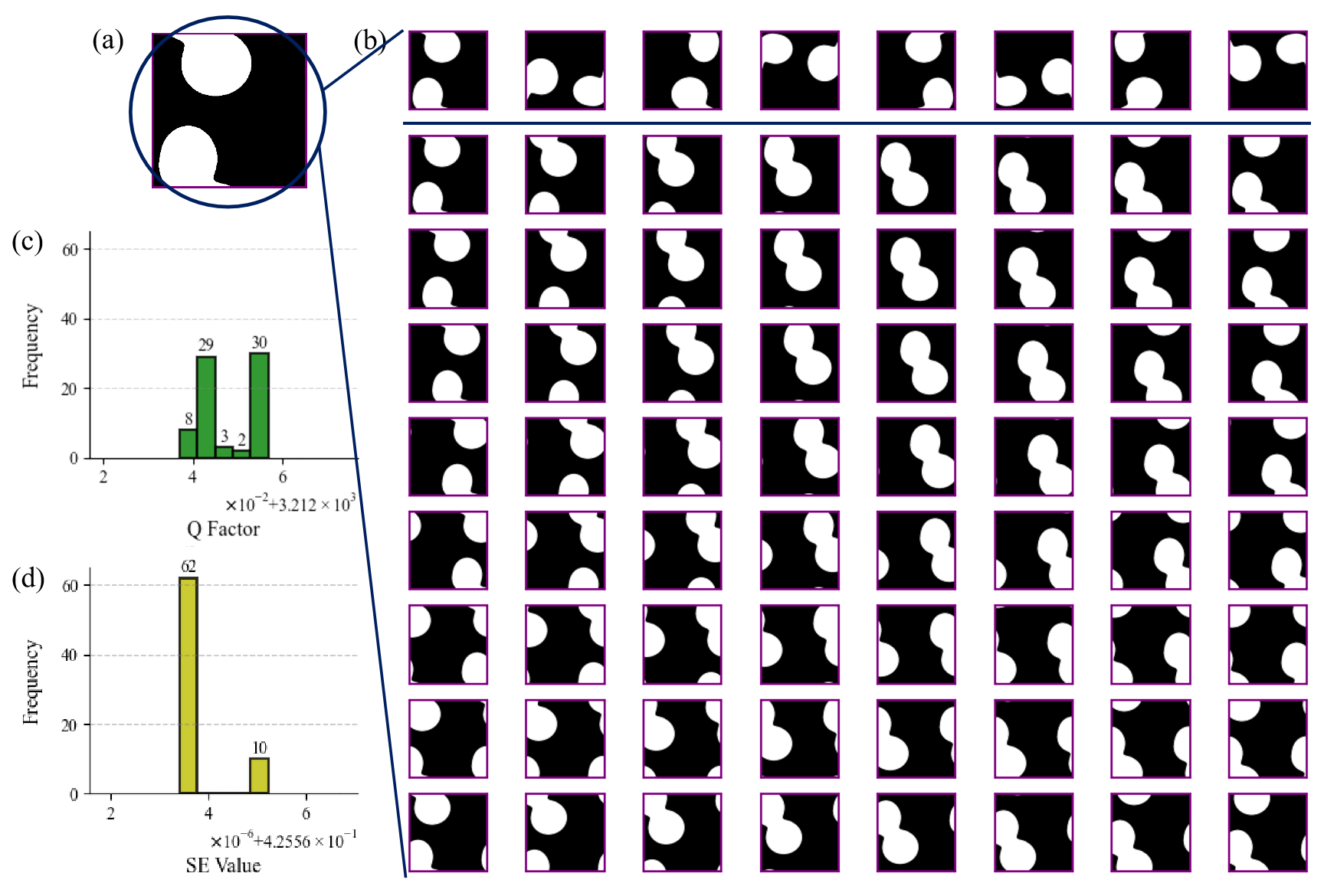}
    \caption{\textbf{Flip-Rotate-Translate pattern effects simulation.} \textbf{a.} A randomly generated unit cell pattern of the photonic crystal. \textbf{b.} The first row shows eight variants of the lattice structure in subfigure a after flipping and rotation. Rows two to nine show sixty-four additional patterns generated by horizontal or vertical translation of the original structure. \textbf{cd.} Simulation results of Coupled-wave Theory model for the photonic crystal lattice structure shown in b after flipping, rotation, and translation. The left and right histograms show the distributions of simulated Q and SE, respectively.}
    \label{fig:CWT stablity}
\end{figure}

The original dataset used in this work contains 25,000 samples, which we divided into training and test sets at a 4:1 ratio. Since the actual device consists of tens of thousands of lattices arranged in a periodic square matrix, flipping or rotating a single lattice's design pattern theoretically does not affect the final PCSEL properties. Furthermore, we observed that even translating patterns - which alters the edge structures of the periodic square matrix - has negligible impact on overall PCSEL performance. The accompanying Figure \ref{fig:CWT stablity} demonstrates this phenomenon using a randomly generated device pattern subjected to flipping, rotation, and translation operations followed by CWT simulations. The histogram clearly shows highly consistent simulation results across all transformations, with Q-factor variations significantly below 1 and $SE$ variations well under 0.1\%. These results not only confirm the pattern-invariant nature of photonic crystal properties but also validate the reliability of the numerical solution component in our CWT model, particularly its convergence characteristics in the non-analytical portion of the calculations. Therefore, the dataset size can be expanded using the methods mentioned above to improve prediction accuracy. The related research is discussed in Section \ref{sec: Training process}.

\begin{figure}[htbp]
    \centering
    \includegraphics[width=0.9\linewidth, trim=2 2 2 2, clip]{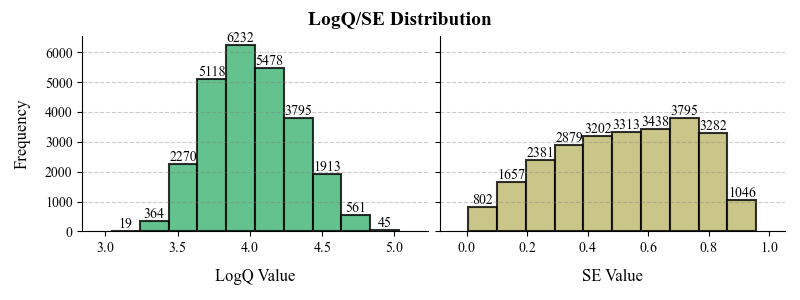}
    \caption{\textbf{Distributions of data.} The histograms show the distributions of logQ and $SE$ for all samples in the original dataset.}
    \label{fig:data distribution}
\end{figure}

\begin{table}[htbp]
    \caption{\textbf{Accuracy vs. optimization targets}}
    \centering
    \renewcommand{\arraystretch}{1.2}
    \begin{tabular}{cc|cc}
        \hline
        \hspace{2em}Target\hspace{2em} & \hspace{2em}$R^2$\hspace{2em} & 
        \hspace{2em}Target\hspace{2em} & \hspace{2em}$R^2$\hspace{2em} \\
        \hline
        \textbf{SE} & \textbf{0.779} & Q & 0.636 \\
        StdSE & 0.775 & StdQ & 0.818 \\
        &  & \textbf{$\log{Q}$} & \textbf{0.909} \\
        &  & Std$\log{Q}$ & 0.865 \\
        \hline
    \end{tabular}
    \footnotetext{This table presents POST's $R^2$ accuracy under different optimization targets. StdQ/StdSE: Linear rescaling of all Q/SE values to [0,1] range; Std$\log{Q}$: Logarithmic transformation of Q values followed by [0,1] rescaling.}
    \label{table:targets}
\end{table}

The choice of prediction targets is also investigated. $SE$ naturally ranges between 0 and 1 with relatively uniform distribution (Figure \ref{fig:data distribution}). Results show that raw $SE$ values achieve the highest prediction accuracy without preprocessing (Table \ref{table:targets}). In contrast, the Q factor can vary dramatically from hundreds to hundreds of thousands. Taking its logarithm yields a more uniform distribution (Figure \ref{fig:data distribution}) and maximizes prediction accuracy (Table \ref{table:targets}). Balanced sample distributions across all value ranges enable the neural network to better distinguish between different photonic crystal designs.

\subsection{POST Backbone Architecture}

\subsubsection{Encoding Module Based on Swin Transformer Block}
To enhance the extraction of feature representations from the single-channel input image \(I\), we used a Swin Transformer Block-based encoder. The resulting multi-dimensional representation \( \mathcal{Z}_4 \) is subsequently passed through an output layer to produce the final prediction results. The overall formulation of the encoding process is expressed as:

\begin{equation}
  \mathrm{SwinTransformer}(I) = \mathcal{Z}_4, \quad I \in \mathbb{R}^{H \times W \times 1}
\end{equation}

\begin{figure}[ht]
  \centering
  \includegraphics[width=0.8\textwidth]{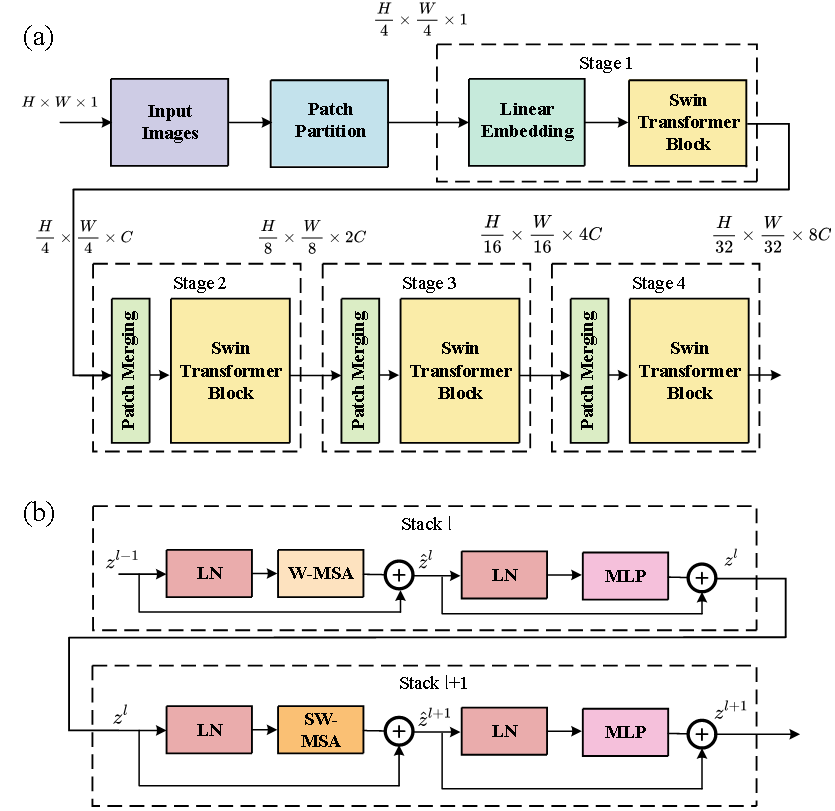}
  \caption{\textbf{POST Network Structure}. Graph \textbf{a} shows the architecture of the Swin Transformer (SwinT) encoder. The input single-channel image is partitioned into patches, linearly embedded, then processed through four hierarchical stages to produce the final multi-dimensional representation \(\mathcal{Z}_4\). Graph \textbf{b} displays two consecutive Swin Transformer stacks. W-MSA and SW-MSA refer to multi-head self attention modules with regular and shifted window configurations.}
  \label{fig:overall_architecture}
\end{figure}

In accordance with the requirements of our task, we adopted the architecture illustrated in Figure~\ref{fig:overall_architecture}(a). The input image is first divided into non-overlapping patches, each of which is treated as an individual token. These tokens are subsequently projected to a predefined feature dimension \(C\) through a linear embedding layer, enabling the subsequent transformer layers to more effectively capture relevant features.

In the proposed architecture, the input tokens are initially passed through a linear embedding layer and subsequently processed by a sequence of modified self-attention modules, known as Swin Transformer Block. These modules operate within non-overlapping local windows to capture spatially localized features while maintaining the number of tokens. We define this initial structure—comprising the linear embedding and Swin Transformer Block—as Stage 1, which serves as the basis for subsequent hierarchical processing. \par

To generate more efficient hierarchical representations, SwinT incorporates a patch merging mechanism\cite{liu2021swin}. As the network deepens, the number of tokens is progressively reduced through these patch merging layers, thereby improving the model’s computational efficiency. For instance, in Stage 2, features from each group of $2 \times 2$ neighboring patches are concatenated into a single vector of dimension $4C$, followed by a linear layer that projects it down to $2C$ dimensions. This operation reduces the tokens count to one-fourth of the previous stage, effectively fusing local information for subsequent processing. The resulting representation, with a resolution of $\frac{H}{8} \times \frac{W}{8}$, is then processed by additional Swin Transformer Block.\par
This design is extended to Stage 3 and Stage 4, where the resolutions are further reduced to $\frac{H}{16} \times \frac{W}{16}$ and $\frac{H}{32} \times \frac{W}{32}$ respectively, resulting in progressively abstract multi-dimensional representations. The complete formulation of the process is provided in Supplementary information.


\subsubsection{Swin Transformer Block}

SwinT enhances the model’s capacity for global information integration by incorporating a shifted window mechanism \cite{liu2021swin} into its architecture and constructing the Swin Transformer Block, as illustrated in Figure~\ref{fig:overall_architecture}(a). This module is a modification of the standard Transformer block, in which the conventional multi-head self-attention (MSA) is replaced with localized attention mechanisms operating within regular and shifted windows—referred to as Window-based MSA (W-MSA) and Shifted Window MSA (SW-MSA), respectively. This design facilitates cross-window information while significantly reducing computational complexity.
The forward propagation of two consecutive Swin Transformer stacks is depicted in Figure~\ref{fig:overall_architecture}(b) and the detailed algorithmic procedures are provided in Supplementary information.

Specifically, each Swin Transformer Block in the architecture comprises two primary components: a window-based self-attention module (either W-MSA or SW-MSA), and a two-layer multilayer perceptron (MLP) equipped with the GELU activation function. Layer normalization (LN) is applied before each submodule, while residual connections are employed following each submodule to enhance training stability in deep networks.

\subsection{Training Process}\label{sec: Training process}


\begin{table}[htbp]
    \caption{\textbf{Data augmentation boosts accuracy}}
    \centering
    \small 
    \setlength{\tabcolsep}{3pt} 
    \renewcommand{\arraystretch}{1.1} 
    \begin{tabular}{ccccc}
        \hline
        Rotations \& Flips & Translation(s) & $R^2$ (SE) & $R^2$ ($\log{Q}$) & Speed (s/epoch) \\
        \hline
        no  & 1\footnotemark[1] & 0.473 & 0.649 & 13 \\
        yes & 1       & 0.654 & 0.845 & 102 \\
        yes & 2       & 0.665 & 0.855 & 407 \\
        \textbf{yes} & \textbf{3} & \textbf{0.779} & \textbf{0.909} & \textbf{919} \\
        yes & 4       & 0.650 & 0.851 & 1623 \\
        yes & 5       & 0.775 & 0.913 & 2552 \\
        yes & 6       & 0.784 & 0.907 & 3678 \\
        \hline
    \end{tabular}
    \footnotetext{The table compares POST's performance across different data augmentation strategies. The combination of 4 flips, 4 rotations, and 6 translations yields the best $SE$ accuracy ($R^2$ = 0.784), while using 5 translations achieves the highest $\log{Q}$ accuracy ($R^2$ = 0.913).}
    \footnotetext[1]{Translation=1 means no additional translation operation is applied.}
    \label{table:data expansion}
\end{table}

Data augmentation through geometric transformations of device patterns proves essential in preprocessing. A single simulation sample can generate 144 valid training samples via 4 flips, 4 rotations, and 3 translations, effectively expanding the original 20,000-sample training set to nearly 3 million samples while significantly reducing additional dataset generation costs. This operation substantially enhances prediction accuracy because neural networks struggle to inherently learn the rotational, reflectional, and translational symmetry properties of PCSEL photonic crystal layers from individual unit cells alone.

To validate this approach, we systematically evaluated POST model performance with different augmentation strategies (Table \ref{table:data expansion}). The first two rows demonstrate improved Q-factor and $SE$ prediction accuracy through rotation and flipping. While increasing translation iterations revealed oscillating accuracy patterns – with odd-numbered translations extracting more meaningful features – excessive translations may cause $R^2$ degradation due to premature overfitting. Notably, since training set size grows quadratically with translation iterations, we ultimately selected 4 flips, 4 rotations, and 3 translations to optimally balance training efficiency and model precision. It should be noted that "1 translation" here means no additional translation operation is performed.

For the loss function, we employed the conventional Mean Squared Error (MSE) (Equation \ref{eq:MSE}), while adopting the $R^2$ metric (Equation \ref{eq:R2}) as our primary evaluation strategy to intuitively assess prediction accuracy and compare model performance between different photonic crystal properties. The $R^2$ metric provides an interpretable scale where: $R^2$=0 indicates that the model's predictions are no better than simply using the mean of the property values, while $R^2$=1 represents perfect prediction accuracy. Higher $R^2$ values correspond to better predictive performance.

\begin{equation}
\text{MSE} = \frac{1}{n} \sum_{i=1}^{n} (y_i - \hat{y}_i)^2 
\label{eq:MSE}
\end{equation}

\begin{equation}
R^2 = 1 - \frac{\sum_{i=1}^{n} (y_i - \hat{y}_i)^2}{\sum_{i=1}^{n} (y_i - \bar{y})^2} 
= 1 - \frac{\text{MSE}}{C^*}, 
\label{eq:R2}
\end{equation}
where $ C^* = \frac{1}{n}\sum_{i=1}^{n} (y_i - \bar{y})^2 $ is a dataset-dependent constant.



\subsection{Model Performance Analysis}

\subsubsection{Comparison to Other Neuron Networks}

\begin{table}[htbp]
    \caption{\textbf{Comparison in Accuracy and Speed.}}
    \centering
    \renewcommand{\arraystretch}{1.2}
    \begin{tabular}{cccc}
        \hline
        \multicolumn{4}{c}{\textbf{$\log{Q}$}} \\
        \hline
        Neuron Network & Train Speed (s) & 
        Predict Speed (s) & $R^2$ of Test Set \\
        \hline
        FCNN & 95 & 0.15 & 0.749 \\
        CNN & 95 & 0.18 & 0.706 \\
        AlexNet \cite{krizhevsky2012imagenet} & 492 & 0.51 & 0.817 \\
        DeiT-Ti \cite{touvron2021training} & 433 & 0.52 & 0.869 \\
        CaiT-S24 \cite{touvron2021going} & 1630 & 1.69 & 0.883 \\
        ConViT-Ti \cite{d2021convit} & 743 & 0.92 & 0.882 \\
        LeViT-128s \cite{graham2021levit} & 687 & 0.84 & 0.881 \\
        \textbf{POST (this work)} & \textbf{918} & \textbf{1.08} & \textbf{0.909} \\
        \hline
        \\
        \hline
        \multicolumn{4}{c}{SE} \\
        \hline
        Neuron Network & Train Speed (s) & 
        Predict Speed (s) & $R^2$ of Test Set \\
        \hline
        FCNN & 80 & 0.16 & 0.570 \\
        CNN & 95 & 0.18 & 0.545 \\
        AlexNet \cite{krizhevsky2012imagenet} & 492 & 0.51 & 0.667 \\
        DeiT-Ti \cite{touvron2021training} & 432 & 0.52 & 0.749 \\
        CaiT-S24 \cite{touvron2021going} & 1621 & 1.70 & 0.763 \\
        ConViT-Ti \cite{d2021convit} & 743 & 0.91 & 0.731 \\
        LeViT-128s \cite{graham2021levit} & 683 & 0.84 & 0.722 \\
        \textbf{POST (this work)} & \textbf{920} & \textbf{1.09} & \textbf{0.779} \\
        \hline
    \end{tabular}
    \footnotetext{The tables compare the prediction performance of different neural networks for two key PCSEL properties ($\log{Q}$ and $SE$), evaluating three critical metrics: training speed (seconds/epoch), prediction throughput (seconds/$5\times10^3$ samples), and test set $R^2$ scores.}
    \label{table:NNs}
\end{table}

A performance comparison of multiple existing neural networks is provided in Table \ref{table:NNs} \cite{krizhevsky2012imagenet, touvron2021training, touvron2021going, d2021convit, graham2021levit}. The evaluation includes not only traditional deep learning models (e.g., FCNN, CNN, and AlexNet) but also various ViT architectures, all tested on the same dataset. The table reveals that POST achieves the highest accuracy (highest $R^2$) in predicting both $\log{Q}$ and $SE$. For $\log{Q}$ prediction, POST attains an $R^2$ of 0.909, outperforming the second-tier models CaiT (0.883), ConViT (0.882), and LeViT (0.881). Similarly, in $SE$ prediction, POST leads with an $R^2$ of 0.779, surpassing CaiT (0.763). These results indicate that POST's architectural design excels at processing smaller-scale images and more accurately captures the physical features of photonic crystal lattice structures, whereas most ViT models exhibit significant advantages only when handling images larger than 100 pixels. Nevertheless, the second-tier ViT models still surpass the milestone model AlexNet in prediction accuracy, highlighting the overall superiority of ViT architectures.

It can be observed that traditional small-scale convolutional neural networks (CNNs) exhibit slightly lower predictive performance than fully connected neural networks (FCNNs). This is attributed to the fact that small CNNs, relying on convolutional mechanisms, can only identify local correlations between pixels and their neighbors, failing to extract global information. In contrast, analyzing the optical field of photonic crystals requires consideration of the entire unit cell structure.


While POST requires greater computational resources than lightweight models, its speed of $>1\times$$10^8$ samples/day and superior precision make it fully capable of supporting more optimization applications.

\begin{figure}[htbp]
    \centering
    \includegraphics[width=0.75\linewidth, trim=2 2 2 4, clip]{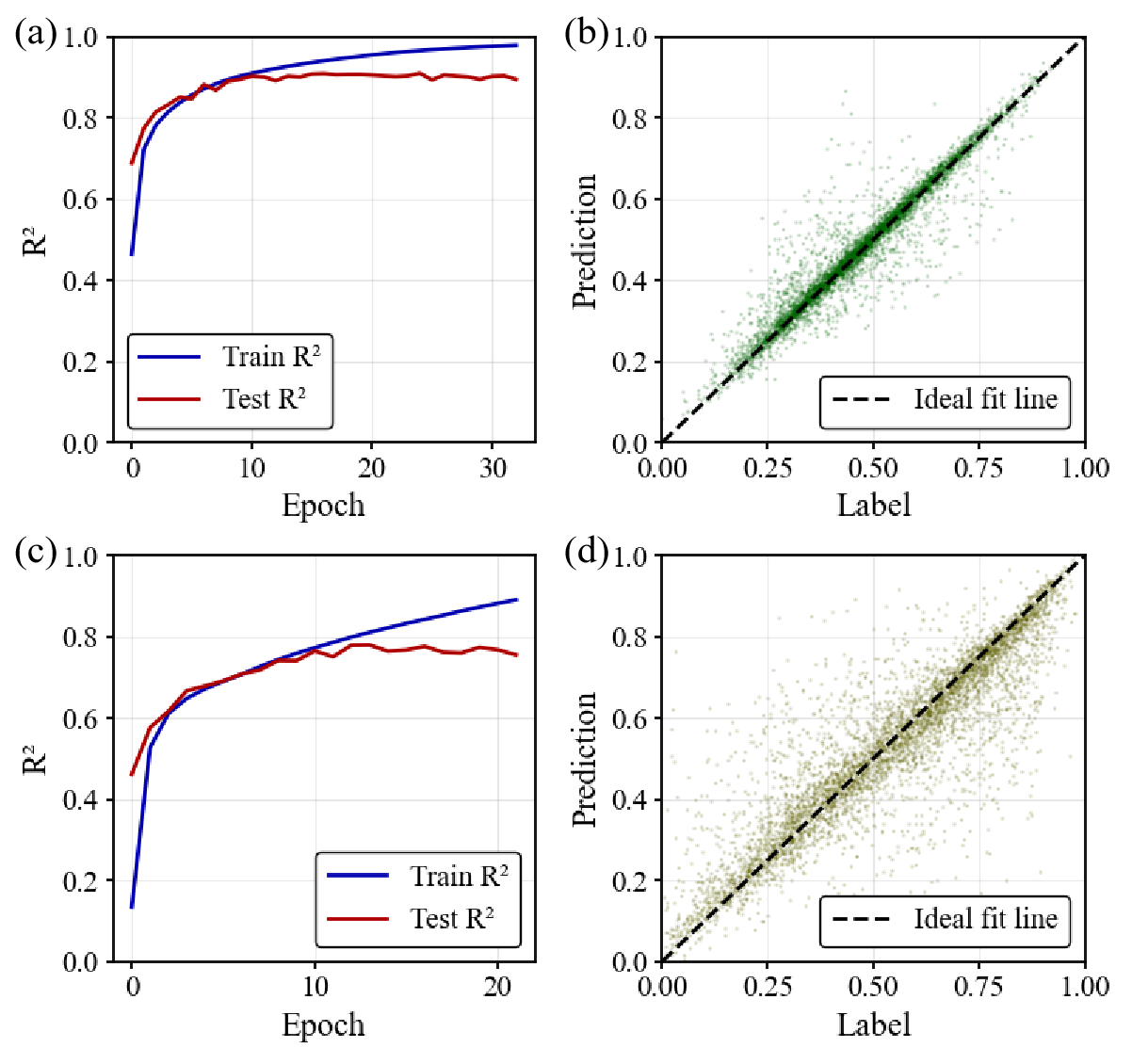}
    \caption{\textbf{Training dynamics and predictive performance.} Graphs \textbf{a} and \textbf{c} show the training curves of the POST's predictions for $\log{Q}$ and SE, respectively. The blue line (Train $R^2$) and red line (Test $R^2$) indicate the goodness-of-fit of the model on the training set and test set as the training epochs progress. Graphs \textbf{b} and \textbf{d} respectively display scatter plots of the model's predictive performance for $\log{Q}$ and SE. The scatter points compare the model's predicted values with the true values, while the black dashed line represents the ideal fit line used to evaluate prediction accuracy.}
    \label{fig:POSTacc}
\end{figure}

\subsubsection{Comparison between Predictions and Simulations }

POST achieves prediction accuracies ($R^2$) of 0.909 for $\log{Q}$ and 0.779 for $SE$ in PCSEL modeling (shown in Figure \ref{fig:POSTacc}). It typically reaches peak accuracy within fifteen epochs, requiring only about three hours of training time, followed by a slight overfitting trend that leads to a minor decline in test set performance.

The scatter plot visually confirms that POST's prediction accuracy for $\log{Q}$ is significantly higher than for SE. This discrepancy stem from the more complex partial differential numerical solving process involved in $SE$ calculations.

\subsubsection{Accuracy with Limited Training Samples}

\begin{figure}[htbp]
    \centering
    \includegraphics[width=0.9\linewidth, trim=2 2 2 2, clip]{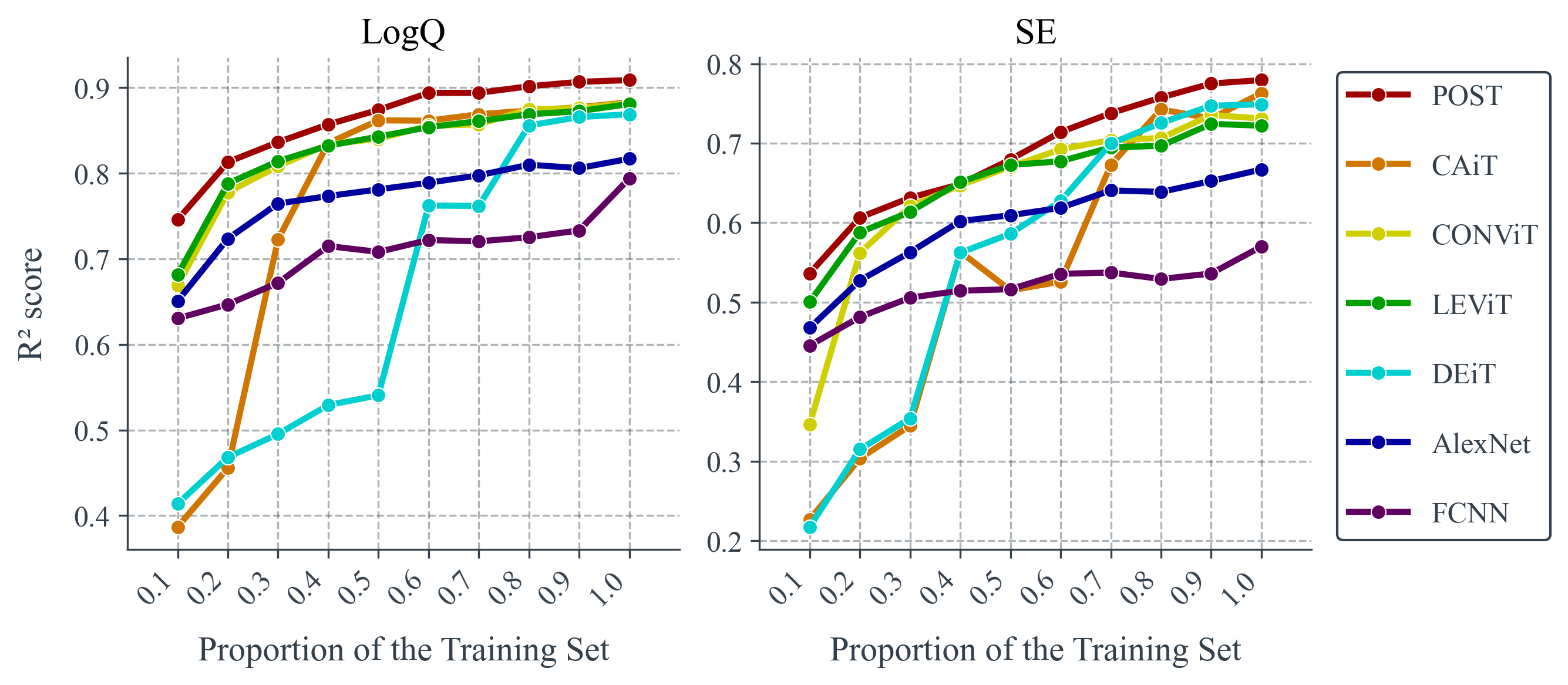}
    \caption{\textbf{Accuracy vs. training set size.} The line chart compares the prediction accuracy (measured by $R^2$) of seven neural networks across varying training set sizes. The horizontal axis represents the proportion of the training set used, relative to the original size of 20,000 samples (80\% of the total 25,000 samples). The reduced training set is then augmented through 4 flips, 4 rotations, and 3 translations before training. The vertical axis shows the corresponding $R^2$ accuracy for each model.}
    \label{fig:R2 v.s. smaller tr set}
\end{figure}

For many simulation software tools, obtaining 25,000 raw data points remains challenging, especially when the input patterns for simulation modules have higher pixel density—simulation time costs increase quadratically. If the simulation input involves 3D structural data, the cost escalates cubically. Therefore, the learning performance of different neural networks under reduced raw data volumes (shown in Figure \ref{fig:R2 v.s. smaller tr set}) is investigated, where the horizontal axis represents the proportion of the new training set relative to the original training set.
 
It can be observed that POST consistently maintains the strongest predictive capability across all dataset sizes and achieves $R^2$ accuracies of $>0.8$ for $\log{Q}$ and $>0.6$ for $SE$ with only 20\% of the original training set. Additionally, we note that POST's prediction accuracy for $\log{Q}$ converges with only about 60\% of the original training set, whereas its accuracy for $SE$ may require a dataset larger than the original training set to converge. This also suggests that predicting the $SE$ parameter is more complex than predicting Q.

Furthermore, Training results with 20\% of the original training set in Figure \ref{fig:R2 v.s. smaller tr set} ($R^2_{SE}$=0.607 and $R^2_{\log{Q}}$=0.813) are worse than those without translation in Table \ref{table:data expansion} ($R^2_{SE}$=0.654 and $R^2_{\log{Q}}$=0.845). However, the former involved three additional translations in both horizontal and vertical directions, resulting in an actual training data volume that was significantly larger than the latter. This suggests that the augmented samples generated through translation carry less additional information compared to entirely new samples.

\subsubsection{Fourier-Space Feature Attribution via SHAP Analysis}

To further investigate whether POST has internalized the physical priors embedded in CWT, we conduct a post hoc interpretability study using SHAP (SHapley Additive exPlanations)\cite{shapley_Notes_1951, lundberg_Unified_2017}. Unlike traditional saliency-based methods that rely on image-space gradients, this analysis evaluates the relative importance of Fourier components in predicting optical properties of PCSEL structures. Specifically, we apply a 2D discrete Fourier transform (DFT) to the photonic crystal unit cell and analyze the real and imaginary parts of selected Fourier coefficients as input features.

For a $32\times32$ dielectric constant distribution representing a photonic crystal pattern, we perform a 2D Fourier transform and extract a subset of its Hermitian-symmetric coefficients according to a triangular masking rule, as illustrated in Figure~\ref{fig:CWT stablity}. This results in 512 unique complex-valued coefficients, each decomposed into real and imaginary parts and concatenated to form a 1024-dimensional input vector. Each index $(m, n)$ corresponds to a Fourier mode $\xi_{m,n} = \xi_{m,n}^R + i\xi_{m,n}^I$.

To quantify the contribution of each Fourier component to the model's predictions, we apply the Permutation SHAP algorithm with 50 background samples and 100 evaluation samples drawn from the test dataset. The prediction function internally recovers the spatial-domain input from each perturbed Fourier vector, enabling seamless compatibility with the original POST architecture trained in the spatial domain.

\begin{figure}[htbp]
    \centering
    \includegraphics[width=0.9\linewidth, trim=2 2 2 2, clip]{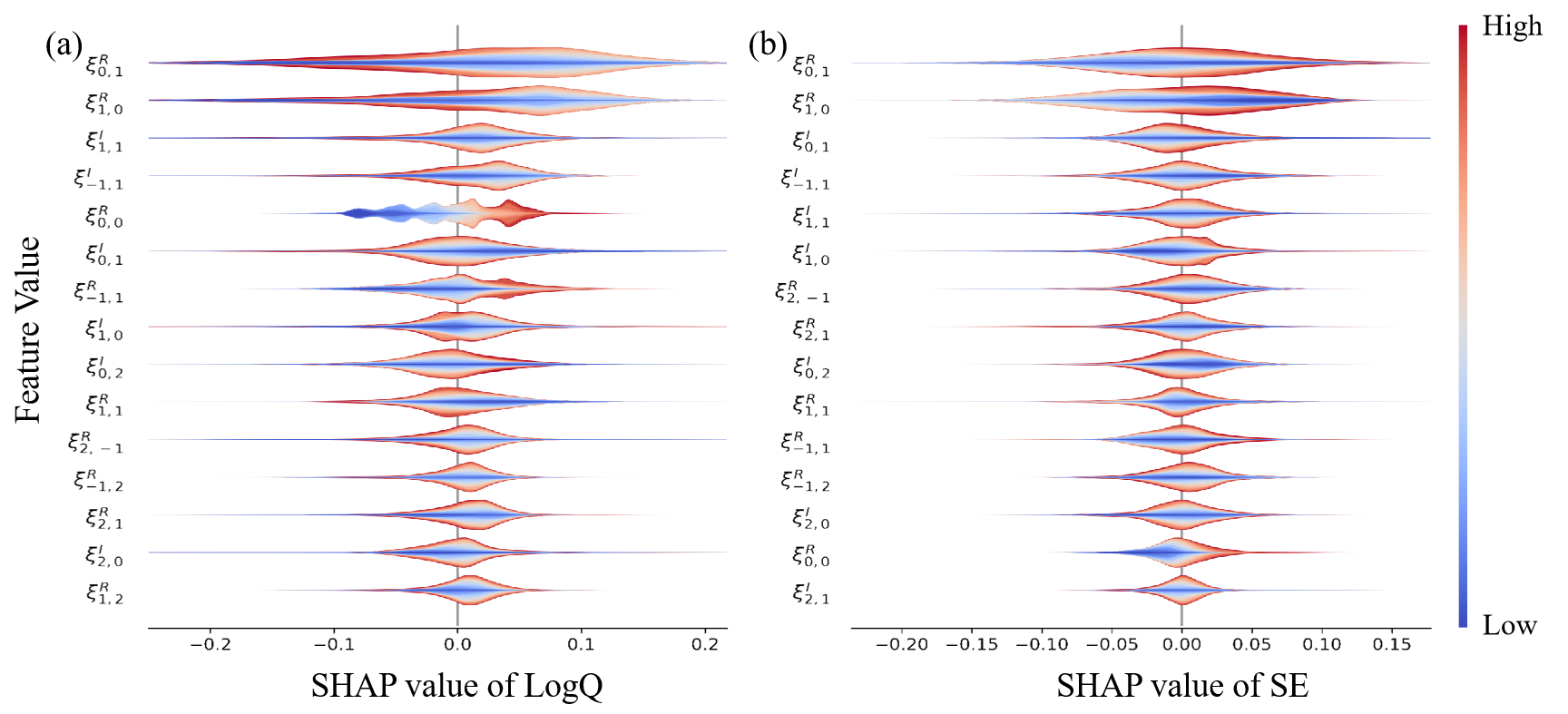}
    \caption{\textbf{Distributions of SHAP values.} Distributions of SHAP values for the top 15 Fourier coefficients contributing to the prediction of \textbf{a} $SE$ and \textbf{b} $\log{Q}$. Each row denotes a distinct coefficient in the Fourier domain, with the x-axis showing its corresponding SHAP value. Color shading indicates the relative magnitude of the coefficient, with red and blue representing high and low values, respectively. The violin plot contours represent the variability and concentration of each coefficient’s contribution. Positive SHAP values suggest a positive influence on the predicted outcome, while negative values imply a suppressive effect.}
    \label{fig:shap_violin}
\end{figure}

Figure~\ref{fig:shap_violin} shows the SHAP violin plots for the top 15 most influential Fourier features for both $SE$ and $\log{Q}$. The most impactful coefficients are concentrated near the center of the Fourier domain, reflecting their critical roles in determining photonic crystal performance.

In particular, $\xi^R_{0,0}$ corresponds to the average refractive index of the photonic crystal layer. A higher value typically indicates a lower filling factor, meaning more high-index material is present. This leads to weaker photonic crystal modulation but stronger waveguiding effect, which enhances vertical optical confinement and contributes positively to the quality factor $Q$.

Coefficients such as $\xi^R_{1,0}$ and $\xi^R_{0,1}$ are directly related to vertical radiation coupling. Deviations of their values from zero increase surface emission efficiency ($SE$) by enhancing out-of-plane leakage. However, since this also introduces greater radiation loss, it tends to reduce $Q$.

In contrast, $\xi^R_{-1,1}$ governs the strength of in-plane two-dimensional diffraction, contributing to lateral distributed feedback. A larger magnitude of this coefficient suggests stronger horizontal coupling, which reinforces resonant feedback and increases $Q$.

Although the top-ranked Fourier features differ slightly between $SE$ and $\log{Q}$, substantial overlap exists in high-impact modes. This indicates that both performance metrics are shaped by a common set of structural features, especially those affecting radiative loss, confinement, and feedback within the photonic crystal.

These findings suggest that POST does not rely merely on low-level image statistics but instead aligns with physical intuition by prioritizing the same global spatial harmonics emphasized by CWT-derived modal coupling analyses. This indicates that the SwinT backbone can implicitly encode physics-relevant representations in the frequency domain.

The SHAP analysis also provides actionable guidance for inverse design tasks. Since only a small subset of Fourier modes contribute significantly to the prediction outcome, optimization algorithms can prioritize this reduced subset of influential coefficients when searching for high-performance PCSEL structures, thus greatly accelerating the convergence of generative design loops.

\subsection{Conclusion}

The authors employed a novel neural network, POST, to predict the photonic crystal simulation results of the CWT model. This approach enables fully automated and highly accurate predictions across single, double, and even triple lattices, as well as various non-circular complex hole structures. It can complete nearly 10,000 predictions in just two seconds, with a mean squared error less than 50\% of previous similar work. Moreover, using only 20\% of the original dataset, it achieves prediction accuracy surpassing prior studies. And SHAP analysis confirms that the model prioritizes physically meaningful Fourier components indicating alignment with CWT theory.

These results demonstrate that POST not only accelerates PCSEL evaluation but also captures key physical principles, making it a promising tool for future AI-assisted photonic device design. The dataset used will be released to support broader research efforts.


\section{Methods}

\subsection{CWT-based Dataset Generation}

The optical properties of photonic crystal surface-emitting lasers (PCSELs) were simulated using a custom three-dimensional coupled-wave theory (3D-CWT) solver implemented in Python. The model accounts for both in-plane diffraction and vertical radiation loss, using up to 441 Fourier harmonics to ensure numerical convergence. For each PCSEL configuration, the surface-emitting efficiency ($SE$) and the logarithmic quality factor ($\log Q$) were computed. The refractive indices used in the simulations are listed in Table~\ref{tab:epitaxy-structure}. The photonic crystal layer was discretized on a $32 \times 32$ grid, and simulations were repeated for over 25,000 unique geometries.

\subsection{Neural Network Model: Photonic Swin Transformer (POST)}

The POST model is based on the Swin Transformer architecture and was implemented using PyTorch. The model takes a single-channel $32 \times 32$ real-space dielectric pattern as input and passes it through four hierarchical self-attention stages. For training, we used an Adam optimizer with a learning rate of $10^{-4}$ and batch size of 64. Separate models were trained for $SE$ and $\log Q$ using mean squared error loss. Training and evaluation were performed on a single NVIDIA RTX 3070 GPU.

\subsection{SHAP Analysis in the Fourier Domain}

To interpret the model’s prediction behavior, we used the PermutationExplainer from the SHAP Python library \cite{lundberg_Unified_2017}. SHAP values were computed in the Fourier domain using 50 background samples and 100 test samples. The model’s internal prediction logic includes inverse Fourier recovery to spatial domain before inference. Violin plots of SHAP values (Fig.~\ref{fig:shap_violin}) were used to identify the most influential modes for $SE$ and $\log Q$.


\section*{Data availability}

Data is provided within the manuscript or supplementary information files.

\backmatter

\newpage










\noindent

    \section*{Acknowledgements}

    This research was supported by the National Natural Science Foundation of China (NSFC) under Grant No. 62174144, the Shenzhen Science and Technology Program under Grants No. JCYJ20220818102214030, No. KJZD20230923115114027, No. JSGG20220831093602004, No. KJZD20240903095602004, the Guangdong Key Laboratory of Optoelectronic Materials and Chips under Grant No. 2022KSYS014, the Shenzhen Key Laboratory Project under Grant No. ZDSYS201603311644527, the Longgang Key Laboratory Project under Grants No. ZSYS2017003 and No. LGKCZSYS2018000015, and the Innovation Program for Quantum Science and Technology (Grant No. 2021ZD0300701), Hefei National Laboratory, Hefei 230088, China, Materials Characterization and Preparation Center, The Chinese University of Hong Kong, Shenzhen.
    
    The authors would thank Ms. Floria Chen for their mentorship and revision tips.

    \section*{Author contributions}

     Q.X. and H.H. conceived the experiment. Q.X. conducted the majority of the manuscript writing and the experiment. H.H., Q.X. and K.S. conducted the experiment, analyzed the results and conducted the correspond part of manuscript. Z.Z. provided guidance and funding support. All authors reviewed the manuscript.

    \section*{Competing interests} The authors declare no conflicts of interest.
    
    \section*{Additional information}

    \subsection*{Supplementary information} 

    The supplementary information includes the dataset and relevant code, provided in the attached files \texttt{supplementary\_information.pdf} and \texttt{ml\_data.zip}.
    
    \subsection*{Correspondence}
    Correspondence and requests for materials should be addressed to Zhaoyu Zhang.

\bigskip

\bibliography{b.sn-bibliography,haihuang_ref}

\end{document}